\documentclass[a4paper,12 pt]{article}

\usepackage{amsmath} 
\usepackage{amsthm}
\usepackage{amssymb}
\usepackage{amsopn}

\begin{document}

\title{Economics needs a scientific revolution}
\author{JP Bouchaud, Science \& Finance, Capital Fund Management, \\
6 Bd Haussmann, 75009 Paris, France}
\maketitle
\begin{abstract}
I argue that the current financial crisis highlights the crucial need of a change of mindset in economics and
financial engineering, that should move away from dogmatic axioms and focus more on data, orders of magnitudes, 
and plausible, albeit non rigorous, arguments. An edited version of this essay appeared in Nature.
\end{abstract}

Compared to physics, it seems fair to say that the quantitative success of the economic sciences is disappointing. Rockets fly to the moon, energy is extracted from minute changes of atomic mass without major havoc, global positioning satellites help millions of people to find their way home. What is the flagship achievement of economics, apart from its recurrent inability to predict and avert crises, including the current worldwide credit crunch?  

Why is this so? Of course, modelling the madness of people is more difficult than the motion of planets, as Newton once said. But the goal here is to describe the behaviour of large populations, for which statistical regularities should emerge, just as the law of ideal gases emerge from the incredibly chaotic motion of individual molecules. To me, the crucial difference between physical sciences and economics or financial mathematics is rather the relative role of concepts, equations and empirical data. Classical economics is built on very strong assumptions that quickly become axioms: the rationality of economic agents, the 'invisible hand' and market efficiency, etc. An economist once told me, to my bewilderment: {\it These concepts are so strong that they supersede any empirical observation}. As Robert Nelson argued in his book, {\it Economics as Religion}, the marketplace has been deified. 

Physicists, on the other hand, have learned to be suspicious of axioms and models. If empirical observation is incompatible with the model, the model must be trashed or amended, even if it is conceptually beautiful or mathematically convenient. So many accepted ideas have been proven wrong in the history of physics that physicists have grown to be critical and queasy about their own models. Unfortunately, such healthy scientific revolutions have not yet taken hold in economics, where ideas have solidified into dogmas, that obsess academics as well as decision-makers high up in government agencies and financial institutions. These dogmas are perpetuated through the education system: teaching reality, with all its subtleties and exceptions, is much harder than teaching a beautiful, consistent formula. Students do not question theorems they can use without thinking. Though scores of physicists have been recruited by financial institutions over the last few decades, these physicists seem to have forgotten the methodology of natural sciences as they absorbed and regurgitated the existing economic lore, with no time or liberty to question its foundations.

The supposed omniscience and perfect efficacy of a free market stems from economic work in the 50's and 60's, which with hindsight looks more like propaganda against communism than a plausible scientific description. In reality, markets are not efficient, humans tend to be over-focused in the short-term and blind in the long-term, and errors get amplified through social pressure and herding, ultimately leading to collective irrationality, panic and crashes. Free markets are wild markets. It is foolish to believe that the market can impose its own self-discipline, as was promoted by the US Securities and Exchange Commission in 2004 when it allowed banks to pile up new debt.
 
Reliance on models based on incorrect axioms has clear and large effects. The 'Black-Scholes model' was invented in 1973 to price options assuming that price changes have a Gaussian distribution, i.e. the probability extreme events is deemed negligible. Twenty years ago, unwarranted use of the model to hedge the downfall risk on stock markets spiraled into the October 1987 crash: -23\% drop in a single day, dwarfing the recent hiccups of the markets. Ironically, it is the very use of the crash-free Black-Scholes model that destabilized the market!  This time around, the problem lay in part in the development of structured financial products that packaged sub-prime risk into seemingly respectable high-yield investments. The models used to price them were fundamentally flawed: they underestimated the probability of that multiple borrowers would default on their loans simultaneously. In other words, these models again neglected the very possibility of a global crisis, even as they contributed to triggering one. The financial engineers who developed these models did not even realize that they helped the credit mongers of the financial industry to smuggle their products worldwide -- they were not trained to decipher what their assumptions really meant. 

Surprisingly, there is no framework in classical economics to understand 'wild' markets, even though their existence is so obvious to the layman. Physics, on the other hand, has developed several models allowing one to understand how small perturbations can lead to wild effects. The theory of complexity, developed in the physics literature over the last thirty years, shows that although a system may have an optimum state (such as a state of lowest energy, for example), it is sometimes so hard to identify that the system in fact never settles there. This optimal solution is not only elusive, it is also hyper-fragile to small changes in the environment, and therefore often irrelevant to understanding what is going on. There are good reasons to believe that this complexity paradigm should apply to economic systems in general and financial markets in particular. Simple ideas of equilibrium and linearity (the assumption that small actions produce small effects) do not work. We need to break away from classical economics and develop altogether new tools, as attempted in a still patchy and disorganized way by `behavioral' economists and `econophysicists'. But their fringe endeavour is not taken seriously by mainstream economics. 
 
While work is done to improve models, regulation also needs to improve. Innovations in financial products should be scrutinized, crash tested against extreme scenarios and approved by independent agencies, just as we have done with other potentially lethal industries (chemical, pharmaceutical, aerospace, nuclear energy, etc.). In view of the present mayhem spilling over from the financial industry into every day life, a parallel with these other dangerous human activities seems relevant. 

Most of all, there is a crucial need to change the mindset of those working in economics and financial engineering. They need to move away from what Richard Feynman called {\it Cargo Cult Science}:  a science that follows all the apparent precepts and forms of scientific investigation, while still missing something essential. An overly formal and dogmatic education in the economic sciences and financial mathematics are part of the problem. Economic curriculums need to include more natural science. The prerequisites for more stability in the long run are the development of a more pragmatic and realistic representation of what is going on in financial markets, and to focus on data, which should always supersede perfect equations and aesthetic axioms.

\end{document}